\documentclass[epj]{webofc}
\usepackage[varg]{txfonts}   
\usepackage{graphicx,color} 
\usepackage{bm}       
\usepackage{amsmath}  
\usepackage{amssymb}
\woctitle{21st International Conference on Few-Body Problems in Physics}
\newcommand{\nc}{\newcommand}

\def\beq{\begin{equation}}
\def\eeq{\end{equation}}
\def\beqa{\begin{eqnarray}}
\def\eeqa{\end{eqnarray}}

\nc{\bP}{{\bf P}}
\nc{\bk}{{\bf k}}
\nc{\bp}{{\bf p}}
\nc{\bpp}{{\bf p}'}

\nc{\bK}{{\bf K}}

\nc{\bq}{{\bf k}}
\nc{\bpjk}{{\bf p}}
\nc{\bqi}{{\bf k}}
\nc{\bpki}{{\bf p}}
\nc{\bqj}{{\bf k}}
\nc{\bpij}{{\bf p}}
\nc{\bqk}{{\bf k}}

\nc{\bki}{{\bf k}}
\nc{\bkj}{{\bf k}}
\nc{\bkk}{{\bf k}}
\nc{\pjk}{p}
\nc{\qi}{k}
\nc{\pki}{p}
\nc{\qj}{k}
\nc{\pij}{p}
\nc{\qk}{k}
\nc{\ki}{k}
\nc{\kj}{k}
\nc{\kk}{k}
\nc{\bera}{\langle}
\nc{\ket}{\rangle}
\nc{\bpinr}{\boldsymbol \pi_{nr}}
\nc{\btpinr}{\tilde{\boldsymbol \pi}_{nrx}}

\nc{\bpi}{\boldsymbol \pi}
\nc{\btpi}{\tilde{\boldsymbol \pi}}

\nc{\pinr}{ \pi_{nr}}
\nc{\tpinr}{\tilde{\pi}_{nr}}
\nc{\bQ}{\bp}
\nc{\bv}{{\bf v}}

\nc{\tpi}{\tilde{\pi}}
\nc{\om}{\omega_m}

\begin{document}
\title{
The Relativistic Three-Body Bound State in Three-Dimensions
}
\author{M. R. Hadizadeh\inst{1}\fnsep\thanks{\email{hadizadm@ohio.edu}} \and
        Ch. Elster\inst{1} \and 
        W.~N.~Polyzou\inst{2} 
}

\institute{
Institute of Nuclear and Particle Physics and Department of Physics and Astronomy, Ohio University, Athens, OH 45701, USA
\and
         Department of Physics and Astronomy, The University of Iowa, Iowa City, IA 52242, USA
          }

\abstract{
 Studying of the relativistic three-body  bound state in a
three-dimensional (3D) approach is a necessary first step in a process
to eventually perform scattering calculations at GeV energies,
where partial-wave  expansions are not useful. To this aim we
recently studied 
relativistic effects in the  
binding energy and for the first time, obtained the
relativistic 3B wave function \cite{Hadizadeh_PRC90}. The
relativistic Faddeev integral equations for the bound state are
formulated in terms of momentum vectors, and 
relativistic invariance is incorporated
within the framework of Poincar\'e invariant quantum mechanics.
}
\maketitle

\section{Introduction and Formalism}
While most three-nucleon calculations utilize partial-wave (PW) representations, 
at laboratory energies above about $\simeq 300$ MeV those become inefficient. 
For applications in the GeV regime one needs to consider vector formulations to implement the
dynamics \cite{Lin_PRC76, Elster_FBS27, Liu_FBS33}. Convergence in the GeV regime has been demonstrated \cite{Lin_PRC76} in a relativistic one channel model using this approach. 
In this work we review the formulation of the 3B bound state problem in vector variables, which demonstrates that the relativistic problem can be treated using a straightforward generalization of methods that have been successfully used in the non-relativistic three-body problem.

In a Faddeev formulation, the relativistic bound state of three particles with mass $m$ interacting with pairwise forces, is described by
\begin{equation} \label{Faddeev}
|\psi \rangle 
= (M_t-M_0)^{-1} \, T(M_t)  \, P \, |\psi \rangle ,
\end{equation}
where $M_t=E_t+3m$ is the $3B$ mass eigenvalue, $M_0$ is the non-interacting $3B$ mass
operator, $T (z) := V + V (z - M)^{-1} V$ is the boosted two-body (2B) $t-$matrix embedded in the
3B Hilbert space, and $P=P_{ij}P + P_{ik}P$ is the permutation operator for three identical 
particles. 
Projecting Eq.~(\ref{Faddeev}) on relativistic momentum states $|\bpjk , \bqi \ket$ leads to the integral equation,
  \begin{equation}
 \hskip-2em \bera \bpjk , \bqi | \psi \ket =
\int d \mathbf{p}' \,  d\mathbf{k}' \,
d \mathbf{p}'' \,  d\mathbf{k}''
{   \delta (\bqi-\bqi' )  \over M_t - M_0 ({p} ,{k})} \, 
T \biggl (\bpjk, \bpjk';M_t- \omega_m (k) \biggr ) \,
\langle \mathbf{p}' ,\mathbf{k}' \vert P \vert \mathbf{p}'' ,\mathbf{k}'' 
\rangle
\,  \bera \bpjk'', \bqi''| \psi \ket .
\end{equation} 
The relativistic basis states contain two Jacobi momentum vectors $\bp$ and $\bk$. The relativistic Jacobi momentum $\bk$ is constructed by boosting the single particle momentum to the 3B rest frame, whereas the momentum $\bp$ is obtained by boosting the momentum $\bk$ to the 2B rest frame.
In the kernel of relativistic Faddeev integral equation one needs the fully off-shell boosted
2B $t-$matrix $T (z)$ which can be calculated from the non-relativistic $t-$matrix $t_{nr}$
by a two-step process, which guarantees that the nonrelativistic and relativistic 2B
observables and CM wave functions remain unchanged. Those steps are
\begin{enumerate}
\item The relativistic right-half-shell 2B $t-$matrix is related to
the non-relativistic right-half-shell by a multiplicative function
$F$ from \cite{Coester_PRC11},
\begin{equation} \label{F-function}
T \left (\bpjk,\bpjk';\sqrt{m_0^2(\bpjk')+\qi^2}+i0^+ \right) = F(\pjk,\pjk',\qi)  \,
t_{nr} \left (\bpjk,\bpjk';\frac{p^{\prime 2}}{m}+i0^+ \right).
\end{equation}
\item The relativistic fully off-shell $t$-matrix is obtained from the relativistic
right-half-shell $t-$matrix by solving a first resolvent type equation,
\begin{equation}
 T(z_j)   =  T(z)  + T(z_j) \, \biggl ( g_0 (z_j) - g_0(z) \biggr ) \, T(z).
\end{equation}
Here $g_0(z)$ is the relativistic two-body propagator, with $z$ being the right-half-shell energy
and $z_j$ the fully-off-shell energy. This procedure  was proposed by Keister and Polyzou \cite{Keister_PRC73} and implemented for the first time by Lin et al. in relativistic Faddeev equations for 3B scattering \cite{Lin_PRC76}.
\end{enumerate}
By following this strategy one can directly obtain the fully off-shell boosted $t-$matrix from nonrelativistic one. 

\noindent
The permutation operator $P$ can be evaluated in relativistic basis states as
\begin{eqnarray}
 \bera \bpjk', \bqi' | P | \bpjk'', \bqi'' \ket  &=& 
   _1\bera \bpjk', \bqi' | \bpki'', \bqj'' \ket_2 +  \, _1\bera \bpjk', \bqi' | \bpij'', \bqk'' \ket _3  \cr
   &=& {\color{black}{N(\bqi',\bqi'')}} 
\Biggl \{ 
\delta^3 \biggl (\bpjk'- \bqi'' - \frac{1}{2} \bqi' C(\bqi',\bqi'') \biggr) \, 
\delta^3 \biggl (\bpjk'' +  \bqi' + \frac{1}{2} \bqi'' C(\bqi'',\bqi') \biggr)  \cr
   &+& \delta^3 \biggl (\bpjk' + \bqi'' + \frac{1}{2} \bqi' C(\bqi',\bqi'') \biggr) \, 
\delta^3 \biggl (\bpjk'' -  \bqi' - \frac{1}{2} \bqi'' C(\bqi'',\bqi') \biggr)
   \Biggr \},
\end{eqnarray}
where
\begin{equation}
\hskip-2em N(\bqi',\bqi'') = {\cal N}^{-1}(-\bqi'-\bqi'',\bqi'') \, {\cal N}^{-1}(-\bqi'-\bqi'',\bqi').
\end{equation}
Here ${\cal N}$ is the square root of the Jacobian 
of the basis change,
\begin{equation}
{\cal N}(\bk_j,\bk_k) = 
\left\vert {\partial  (\bq_j,\bq_k)  \over  
\partial (\bp_{jk}, \bk_{jk}) } \right \vert ^{1/2} =
\left [ {\omega_m(p_{jk}) +\omega_m(p_{jk}) \over 
\omega_m(k_j) +\omega_m(k_k)}
{\omega_m(k_j) \omega_m(k_k) \over 
\omega_m(p_{jk}) \omega_m(p_{jk})} \right ] ^{1/2}  \ne 1.
\label{2.12} 
\end{equation}
The function $C$ is given by
\begin{equation}
\hskip-2em C(\bqi,\bqi') = 1+ \frac{ \omega_m (k')- 
\omega_m(\vert\bqi + \bqi'\vert) }
{ \omega_m(k') + \omega_m(\vert \bqi + \bqi'\vert ) + 
\sqrt{\biggl ( \omega_m(k') 
+ \omega_m (\vert\bqi + \bqi' \vert) \biggr)^2-k^2} }, 
\end{equation}
where $\omega_m (k) = \sqrt{m^2 +k^2}$.
In the large mass limit, i.e. $m \gg \vert \bpjk \vert ,\vert \bki \vert$, the permutation
coefficient $C$ and the Jacobian function $N$ approach the nonrelativistic limit and  are 
equal to one.

After evaluating  the permutation operator and the matrix elements of the fully off-shell
two-body $t-$matrix, the relativistic Faddeev integral equation reads,
\begin{equation} 
\psi  (\bpjk \,, \bqi) =
 \frac{1}{M_t-M_0(p,k)}\int d \bqi' \, N(\bqi,\bqi') \,  T^{sym} 
\biggl(\bpjk,\btpi;M_t-\omega_m (k) \biggr)  \, \psi (\bpi,\bqi') ,
\end{equation}
where $T^{sym} \bigl ( \bpjk,\bpjk';\epsilon \bigr) 
=   T \bigl ( \bpjk,\bpjk';\epsilon \bigr)  +  T\bigl ( -\bpjk,\bpjk';\epsilon \bigr)$ is
symmetrized boosted two-body $t$-matrix and the shifted momentum arguments are given by
\begin{equation} 
\btpi =  \bqi' + \frac{1}{2} C(\bqi,\bqi') \, \bqi  , \quad 
\bpi =   \bqi  + \frac{1}{2} C(\bqi',\bqi) \, \bqi'.
\label{eq.4.28}
\end{equation}

The relativistic and non-relativistic interaction models are defined so that they have
the same two-body center-of-momentum scattering cross sections and
wave functions.  In the three-body space they are fixed by $S$-matrix cluster
properties and Poincar\'e or Galilean invariance, respectively. Differences  
between the relativistic and nonrelativistic
Faddeev equations arise from:
\begin{enumerate}
\item the Jacobian function $N$, representing from the change of the 3B basis states,
\item the coefficient $C$, which appears in the shifted arguments of the momenta,
\item the relation between the relativistic and non-relativistic $t$-matrices defined by function $F$,
\item the relations between the relativistic and non-relativistic free Green's functions.
\end{enumerate}

\section{Results and Discussion}

For our numerical analysis we used two models of a spin-independent Malfliet-Tjon type
potential, MT-V \cite{Elster_FBS27} and MT-Vc \cite{Liu_FBS33}. The model MT-Vc contains an
additional cutoff function of dipole type.  The relativistic Faddeev integral equation is solved 
 using a Lanczos technique \cite{Stadler_PRC44}.
The calculated relativistic and nonrelativistic 3B binding energies for MT-V potential are 
$E_{r}=-7.4825$ (MeV) and $E_{nr} = -7.7382$ (MeV), which indicates a reduction of about
 $(E_r-E_{nr})/E_{nr} =3.3\,\%$ in the 3B binding energy.
This reduction is consistent with the partial wave calculation of Ref. \cite{Gloeckle_PRC33}, truncated to s-wave, with about $2.70 \, \%$ reduction in the 3B binding energy.
The similar behavior can also be seen for MT-Vc potential, where the relativistic effects
leads to about $3.2\,\%$ reduction in 3B binding energy. The obtained nonrelativistic and
relativistic 3B binding energies for MT-Vc potential are $E_{nr} = -7.7177$ (MeV) and
$E_{r}=-7.4732$ (MeV), respectively.

\begin{figure}[ht]
\centering
\sidecaption
\includegraphics[width=9.5cm,clip]{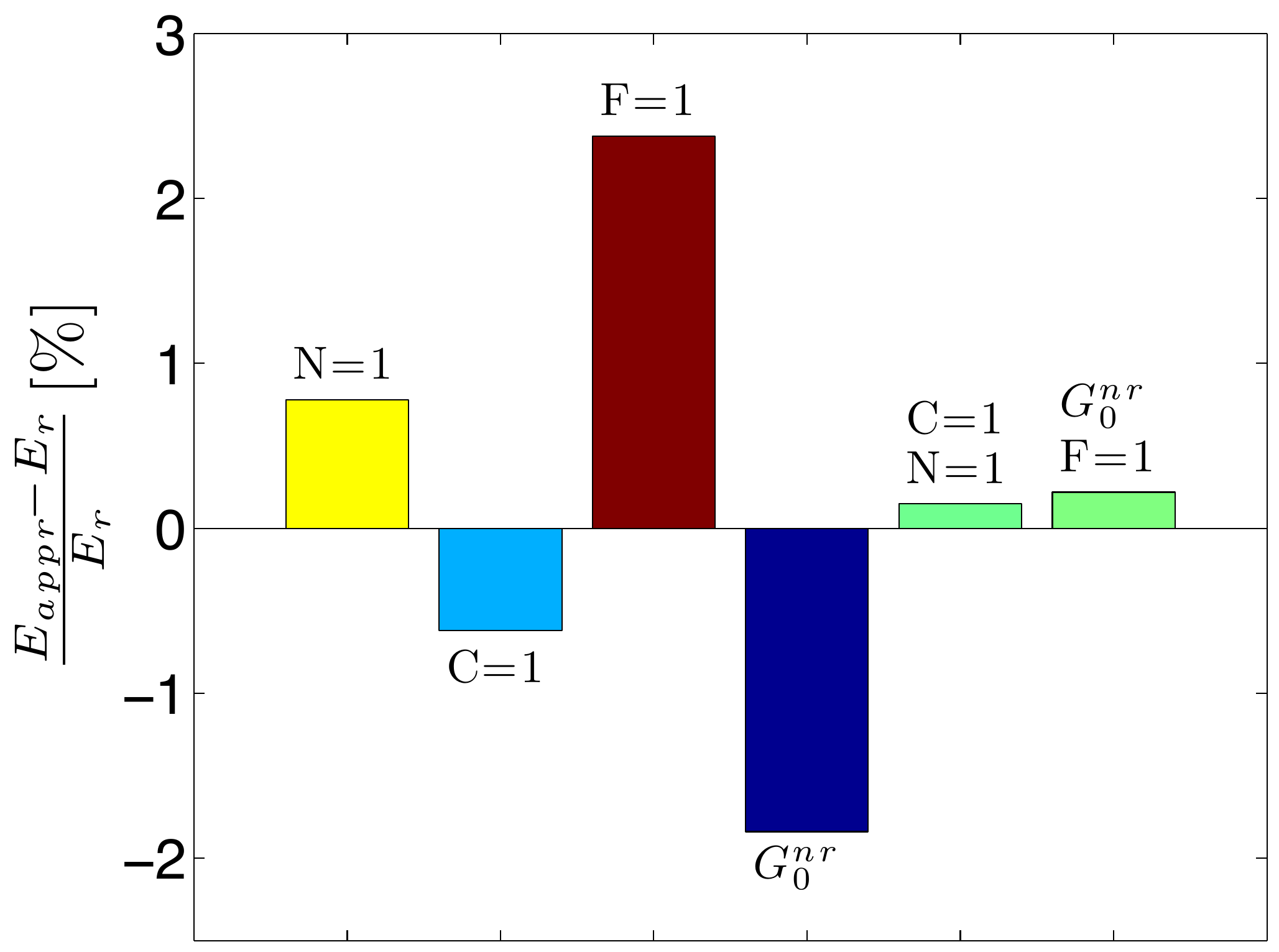}
\caption{Contributions of different relativistic corrections to the 3B
  binding energy of MT-V potential. (1) $N=1$: the Jacobian of the permutation operator (2)
  $C=1$ relativistic Jacobi momenta $\to$ NR Jacobi
  momenta, (3) $F=1$ the relativistic half-shell $t$-matrix $\to$
  NR half-shell $t$-matrix (4) $G_0$: free resolvent $G_0 \to G_{0}^{NR}$.}
\label{fig.bar}
\end{figure}

While our calculations indicate that the overall relativistic effect is small, 
this results from large cancellations in the four relativistic corrections 
mentioned above. 
In Fig. 1, we show the contribution of
each correction to the 3B relativistic binding energy of MT-V potential, where
$E_{appr}$ is the 3B binding energy when one of the ingredients is
replaced by the corresponding non-relativistic quantity.
By setting the Jacobian function $N$ to one in our relativistic calculations, the 3B binding energy has a small increase of about $0.8\,\%$, while setting the permutation coefficient $C$ to one leads to a small reduction of about $0.6\, \%$.
Replacing the relativistic right-half-shell $t-$matrix with the non-relativistic one, i.e.
setting $F=1$ in Eq. (\ref{F-function}),  leads to about $2.4\,\%$ increase in the 3B binding energy.
Finally,  replacing the relativistic free propagator with the nonrelativistic one, the 3B
binding energy is  reduced by about $1.84\,\%$.
By setting both Jacobian functions,  $N$ and $C$, to one in our relativistic calculations,
the net contribution of the dynamic ingredients to the 3B binding energy is a small increase
of about $0.15\, \%$. Similarly, the net contribution of the kinematic ingredients obtained by 
setting $F=1$ and replacing the relativistic free propagator with the nonrelativistic one, 
leads to  an increase in the 3B binding energy of about $0.22\, \%$.

The extension of the formulation of the relativistic
Faddeev integral equations based on realistic nucleon-nucleon (NN) interactions 
is in progress. The input here is an operator
representation of spin-dependent NN interaction in a helicity representation, which
depends on the total spin and the relative momentum in the
NN subsystems. 
Since the 3D approach automatically considers all partial
waves the number of equations is fixed. The successful implementation
in $3N$ bound state will pave the way to $3N$ scattering in the 
few-GeV energy range.

\begin{acknowledgement}
This work was performed under the auspices of the National Science Foundation under Contract No. NSF-PHY-1005587 with Ohio University and Contract No. NSF-PHY-1005501 with the University of Iowa. Partial support was also provided by the U. S. Department of Energy, Office of Science, Office of Nuclear Physics, under Contract No. DE-FG02-93ER40756 with Ohio University, and Contract No. DE-FG02-86ER40286 with the University of Iowa. We thank the Ohio Supercomputer Center (OSC) for the use of their facilities under Grant No. PHS0206.
\end{acknowledgement}

%
%
%

\end{document}